\documentclass[twocolumn,journal]{IEEEtran}
\usepackage[usenames,dvipsnames]{color}
\usepackage{subfigure}
\usepackage{graphicx}
\usepackage{amsmath}
\usepackage{color}
\usepackage{multicol}
\usepackage{amssymb}

\usepackage{amsfonts}%
\usepackage{verbatim}%
\usepackage{enumerate}%
\usepackage{psfrag}
\usepackage{epsfig}
\usepackage{multicol}
\usepackage{setspace}
\usepackage{dsfont}

\usepackage{algorithm}
\usepackage{algorithmic}
\usepackage{cite}
\usepackage{float}

\begin{document}
\title{Optimal Random Access and Random Spectrum Sensing for an Energy Harvesting Cognitive Radio}
\author{ Ahmed El Shafie$^\dagger$, Ahmed Sultan$^*$\\
\small \begin{tabular}{c}
$^\dagger$Wireless Intelligent Networks Center (WINC), Nile University, Giza, Egypt. \\
$^*$Department of Electrical Engineering, Alexandria University, Alexandria, Egypt. \\
\end{tabular}
}
\date{}
\maketitle
\begin{abstract}
We consider a secondary user with energy harvesting capability. We design access schemes for the secondary user which incorporate random spectrum sensing and random access, and which make use of the primary automatic repeat request (ARQ) feedback. The sensing and access probabilities are obtained such that the secondary throughput is maximized under the constraints that both the primary and secondary queues are stable and that the primary queueing delay is kept lower than a specified value needed to guarantee a certain quality of service (QoS) for the primary user. We consider spectrum sensing errors and assume multipacket reception (MPR) capabilities. Numerical results are presented to show the enhanced performance of our proposed system over a random access system, and to demonstrate the benefit of leveraging the primary feedback.
\end{abstract}

\section{Introduction}
Cognitive radio technology provides an efficient means of utilizing the radio spectrum \cite{zhao2007survey}. The basic idea is to allow secondary users to access the spectrum while providing certain guaranteed quality of service (QoS) performance measures for the high-priority licensed primary users (PUs). The secondary user is a battery-powered device in many practical situations and its operation, which involves spectrum sensing and access, is accompanied by energy consumption. Consequently an energy-constrained secondary user must optimize its sensing and access decisions to efficiently utilize the energy at its disposal. An emerging technology for energy-constrained terminals is energy harvesting which allows the terminal to gather energy from its environment. An overview of the different energy harvesting technologies is provided in \cite{survey} and the references therein.

Data transmission by an energy harvesting transmitter with a rechargeable battery has got a lot of attention recently  \cite{lei2009generic,sharma2010optimal,ho2010optimal,yang2010transmission,yang2010optimal,tutuncuoglu2010optimum,pappas2011optimal,krikidis2012stability}. The optimal online policy for controlling admissions into the data buffer
is derived in \cite{lei2009generic} using a dynamic programming framework. In \cite{sharma2010optimal},
energy management policies which stabilize the data queue
are proposed for single-user communication and some delay-optimal properties are derived. Throughput optimal
energy allocation is investigated in \cite{ho2010optimal} for energy harvesting systems in
a time-constrained slotted setting. In \cite{yang2010transmission,yang2010optimal}, minimization
of the transmission completion time is considered in an
energy harvesting system and the optimal solution is obtained
using a geometric framework. In \cite{tutuncuoglu2010optimum}, energy harvesting transmitters with
batteries of finite energy storage capacity are considered and
the problem of throughput maximization by a deadline is
solved for a static channel. The authors of \cite{pappas2011optimal} consider the scenario in which a set of
nodes shares a common channel. The PU has a rechargeable
battery and the secondary user is plugged to a reliable power supply. They obtain the maximum stable throughput region which describes the maximum arrival rates that maintain the stability of the network queues. In \cite{krikidis2012stability}, the authors investigate the effects of network layer
cooperation in a wireless three-node network with energy harvesting
nodes and bursty traffic.

In this work, we develop spectrum sensing and transmission methods for an energy harvesting secondary user. We leverage the primary automatic repeat request (ARQ) feedback for secondary access. Due to the broadcast nature of the wireless channel, this feedback can be overheard and utilized by the secondary node assuming that it is unencrypted. The problem with depending on spectrum sensing only is that sensing does not inform the secondary terminal about its impact on the primary receiver. This issue has induced interest in utilizing the feedback from the primary receiver to the primary transmitter to optimize the secondary transmission
strategies. For instance, in \cite{eswaran2007bits}, the secondary user observes the ARQ feedback from the primary
receiver as it reflects the PU's achieved packet rate. The secondary user's objective is to maximize its throughput while guaranteeing a certain packet rate for the PU. In \cite{lapiccirella2010cognitive}, the authors use a partially observable Markov decision process (POMDP) to optimize the
secondary action on the basis of the spectrum sensing outcome and primary ARQ feedback. Secondary power control based on primary feedback is investigated in \cite{huang2010distributed}. In \cite{levorato2009cognitive}, the optimal transmission policy for the secondary user when the PU adopts a retransmission based error control scheme is investigated. The policy of the secondary user determines how often it transmits according to the retransmission state of the packet being served by the PU.

Our contributions in this paper can be summarized as follows. We investigate the case of a secondary user equipped with an energy harvesting mechanism and a rechargeable battery. We propose a novel access and sensing scheme where the secondary user possibly senses the channel for a certain fraction of the time slot duration and accesses the channel with some access probability that depends on the sensing outcome. The secondary user may access the channel probabilistically without sensing in order to utilize the whole slot duration for transmission. We assume multipacket reception (MPR) capability added to the physical layer of the receiving nodes. We maximize the secondary throughput such that the primary and secondary queues are stable. The optimization problem to get the various probabilities also includes a constraint on the maximum tolerable primary queueing delay. Finally, we compare our system with the random access system. The numerical results show the gains of our proposed systems in terms of the secondary user throughput.

\begin{figure}
  \includegraphics[width=1.05\columnwidth]{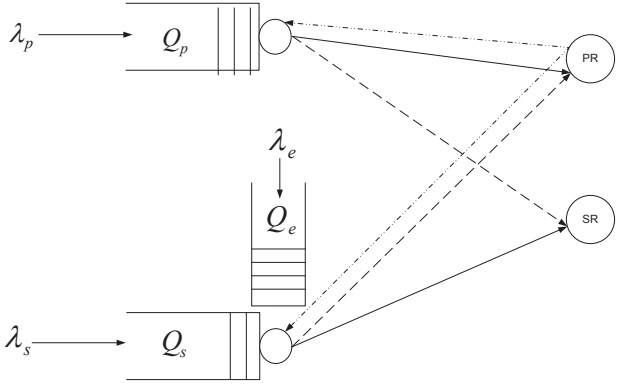}\\
  \caption{Primary and secondary queues and links. The PU has data queue $Q_p$, whereas the secondary terminal has data queue $Q_s$ and energy queue $Q_e$. There is a feedback channel between the primary receiver (PR) and the PU to acknowledge the reception of data packets. This feedback channel is overheard by the secondary transmitter. Both the PR and the secondary receiver (SR) may suffer interference from the other link. }\label{fig1}
\end{figure}

The rest of the paper is organized as follows. In the next section, we discuss secondary access without incorporating the primary feedback. In Section \ref{sec3} we discuss the feedback-based scheme. We provide numerical results in Section \ref{numerical} and conclude the paper in Section \ref{secfinal}.


\section{Secondary Access Without Employing Primary Feedback}\label{sec2}
\subsection{System Model}
We consider the system model shown in Fig. \ref{fig1}. The model consists of one PU and
one secondary user. The channel is slotted in time and
a slot duration equals the packet transmission time. The PU and the secondary user have infinite buffer queues, $Q_p$ and $Q_s$, respectively, to store fixed-length data packets. If a terminal transmits during a time slot, it sends exactly one packet to its receiver. The
arrivals at $Q_p$ and $Q_s$ are independent and identically
distributed (i.i.d.) Bernoulli random variables from slot to slot with means $\lambda_p$ and $\lambda_s$, respectively.

The secondary user has an additional energy queue, $Q_e$, to store harvested energy from the environment. The arrival at the energy queue is also Bernoulli with mean $\lambda_e$ and is independent from arrivals at the other queues. The Bernoulli model is simple, but it captures the random availability of ambient energy sources. It is assumed that the transmission of one data packet consumes one packet of energy. Adequate system operation requires that all the queues are stable. We employ the standard definition of stability for queue as in, e.g., \cite{stabN}, that is, a queue is stable if and only if its probability of being empty does not vanish as time progresses.

Instead of the collision channel model where simultaneous transmission by different terminals leads to sure packet loss, we assume that the receivers have multipacket reception (MPR) capability as in \cite{ghez1988stability,ghez1989optimal,naware2005stability}. This means that transmitted data packets can survive the interference caused by concurrent
transmissions if the received signal to interference
and noise ratio ({\rm SINR}) exceeds the threshold required for successful decoding at the receiver. With MPR capability, the secondary user may use the channel simultaneously with the PU.

The PU accesses the channel whenever
it has a packet to send. The secondary transmitter, given that it has energy, senses the channel or possibly transmits the packet at the head of its queue immediately at the beginning of the time slot without sensing the channel. We explain below why direct transmission can be beneficial for system performance.

The secondary user operation can be summarized as follows.
\begin{itemize}
  \item If the secondary terminal's energy and data queues are not empty, it senses the channel with probability $p_s$
   from the beginning of the time slot for a duration of $\tau$ seconds to detect the possible activity of the PU. If the slot duration is $T$, $\tau < T$.
    \item If the channel is sensed to be free, the secondary transmitter accesses the channel with probability $p_f$. If the PU is detected to be active, it accesses the channel with probability $p_b$.
    \item If at the beginning of the time slot the secondary user decides not to sense the spectrum (which happens with probability $1-p_s$) it immediately decides whether to transmit with probability $p_t$ or to remain idle for the rest of the time slot with probability $1-p_t$.
\end{itemize}
\noindent This means that the transmission duration is $T$ seconds if the secondary user accesses the channel without spectrum sensing and $T-\tau$ seconds if transmission is preceded by a sensing phase. We assume that the energy consumed in spectrum sensing is negligible, whereas data transmission dissipates exactly one unit of energy from $Q_e$.

We study now secondary access in detail to obtain the mean service rates of queues $Q_s$, $Q_e$ and $Q_p$. The meaning of the various relevant symbols are provided in Table \ref{table1}. For the secondary terminal to be served, its energy queue must be nonempty. If the secondary user does not sense the channel, which happens with probability $1-p_s$, it transmits with probability $p_t$. If the PU's queue is empty and, hence, the PU is inactive, secondary transmission is successful with probability $\overline{P}_{{\rm out},0s}$, whose expression as a function of the secondary link parameters, transmission time $T$, and the data packet size is provided in Appendix A. If $Q_p \neq 0$ and the PU is active, the success probability is $\overline{P}_{{\rm out},0s}^{\left({\rm c}\right)}$ (see Appendix B).
\begin{table}
\renewcommand{\arraystretch}{1}
\begin{center}
\begin{tabular}{ c |l  }
    \hline\hline
    $\tau$ & {\footnotesize Sensing duration} \\[5pt]\hline
  $T$ & {\footnotesize Slot duration} \\[5pt]\hline
  $p_s$ & {\footnotesize Probability of sensing the channel} \\[5pt]\hline
   $P_{\rm MD}$ & {\footnotesize Misdetection probability} \\[5pt]\hline
    $P_{\rm FA}$ & {\footnotesize False alarm probability} \\[5pt]\hline
  $p_t$ & {\footnotesize Probability of direct channel access if the channel is } \\
   & {\footnotesize not sensed} \\[5pt]\hline
  $p_f$ & {\footnotesize Probability of channel access if the channel is sensed} \\
   & {\footnotesize to be free} \\[5pt]\hline\\
    $p_b$ & {\footnotesize Probability of channel access if the channel is sensed} \\
   & {\footnotesize to be busy} \\[5pt]\hline\\
  $\overline{P}_{{\rm out},p}$ & {\footnotesize Probability of successful primary transmission to the } \\
   & {\footnotesize primary receiver if the secondary terminal is silent}\\[5pt]\hline
   $\overline{P}_{{\rm out},p}^{\left({\rm c}\right)}$ & {\footnotesize Probability of successful primary transmission to the } \\
   & {\footnotesize primary receiver with concurrent secondary transmission}\\[5pt]\hline
 $\overline{P}_{{\rm out},0s}$ & {\footnotesize Probability of successful secondary transmission if the PU} \\
   & {\footnotesize is silent and transmission occurs over $T$ seconds}\\[5pt]\hline
   $\overline{P}_{{\rm out},1s}$ & {\footnotesize Probability of successful secondary transmission if the PU} \\
   & {\footnotesize is silent and transmission occurs over $T-\tau$ seconds}\\[5pt]\hline
   $\overline{P}_{{\rm out},0s}^{\left({\rm c}\right)}$ & {\footnotesize Probability of successful secondary transmission if the PU} \\
   & {\footnotesize is active and transmission occurs over $T$ seconds}\\[5pt]\hline
   $\overline{P}_{{\rm out},1s}^{\left({\rm c}\right)}$ & {\footnotesize Probability of successful secondary transmission if the PU} \\
   & {\footnotesize is active and transmission occurs over $T-\tau$ seconds}\\[5pt]\hline
\end{tabular}
\caption{List of symbols involved in the queues' mean service rates.}
\label{table1}
\end{center}
\end{table}

If the secondary user decides to sense the channel, there are four possibilities depending on the sensing outcome and the state of the primary queue. If the PU is sensed to be free, secondary transmission takes place with probability $p_f$. This takes place with probability $1-P_{\rm FA}$ if the PU is actually silent. In this case, the probability of successful secondary transmission is $\overline{P}_{{\rm out},1s}$, which is lower than $\overline{P}_{{\rm out},0s}$ as proven in Appendix C. On the other hand, if the PU is on, the probability of detecting the channel to be free is $P_{\rm MD}$ and the probability of successful secondary transmission is $\overline{P}_{{\rm out},1s}^{\left({\rm c}\right)}$. If the channel is sensed to be busy, the secondary terminal transmits with probability $p_b$. Sensing the PU to be active occurs with probability $P_{\rm FA}$ if the PU is actually inactive, or with probability $1-P_{\rm MD}$ if the PU is actively transmitting. The probability of successful secondary transmission is $\overline{P}_{{\rm out},1s}$ when the PU is silent and $\overline{P}_{{\rm out},1s}^{\left({\rm c}\right)}$ when the PU is active. Given these possibilities, we can write the following expression for the mean secondary service rate.
\begin{equation}
\begin{split}
\mu_s&=\left(1-p_s\right)p_t{\rm Pr}\{Q_p =0,Q_e \ne 0\} {\overline{P}}_{{\rm out},0s}
\\&+\left(1-p_s\right)p_t{\rm Pr}\{Q_p \ne 0,Q_e \ne 0\} {\overline{P}}_{{\rm out},0s}^{\left({\rm c}\right)}
\\& +p_s p_f {\rm Pr}\{Q_p =0,Q_e \ne 0\} \left(1-P_{\rm FA}\right){\overline{P}}_{{\rm out},1s}
\\&+p_s p_f {\rm Pr}\{Q_p \ne0,Q_e \ne 0\} P_{\rm MD} {\overline{P}}_{{\rm out},1s}^{\left({\rm c}\right)}
\\&+p_s p_b {\rm Pr}\{Q_p =0,Q_e \ne 0\}P_{\rm FA} {\overline{P}}_{{\rm out},1s}
\\&+p_s p_b  {\rm Pr}\{Q_p \ne0,Q_e \ne 0\} \left(1-P_{\rm MD}\right) {\overline{P}}_{{\rm out},1s}^{\left({\rm c}\right)}
\end{split}
\end{equation}

\noindent Based on the above analysis, it can be shown that the mean service rate of the energy queue is
\begin{equation}
\begin{split}
\mu_e&=\left(1-p_s \right)p_t{\rm Pr}\{Q_s \ne 0\}\\& +p_s p_f \bigg(P_{\rm MD} {\rm Pr}\{Q_s \ne 0,Q_p\ne 0\}\\&+(1-P_{\rm FA}){\rm Pr}\{Q_s \ne 0,Q_p=0\}\bigg) \\& +p_s p_b \bigg(P_{\rm FA} {\rm Pr}\{Q_s \ne 0,Q_p=0\}\\&+(1-P_{\rm MD}){\rm Pr}\{Q_s \ne 0,Q_p\ne 0\}\bigg)
\end{split}
\end{equation}

A packet from the primary queue can be served in either one of the following events. If the secondary user is silent because either of its data queue or energy queue is empty, the primary transmission is successful with probability $\overline{P}_{{\rm out},p}$. If both secondary queues are nonempty, secondary operation proceeds as explained above. In all cases, if the secondary user does not access the channel, the probability of successful primary transmission is $\overline{P}_{{\rm out},p}$, else it is $\overline{P}_{{\rm out},p}^{\left({\rm c}\right)}$. Therefore,
\begin{equation}
\begin{split}
\mu_p&=\bigg(1-{\rm Pr}\{Q_s \ne0,Q_e \ne 0\}\bigg)\overline{P}_{{\rm out},p}
\\&+{\rm Pr}\{Q_s \ne0,Q_e \ne 0\}\Bigg(
\\&\left(1-p_s\right)\bigg[p_t \overline{P}_{{\rm out},p}^{\left({\rm c}\right)}+\left(1-p_t\right)\overline{P}_{{\rm out},p}\bigg]
\\&+p_s P_{\rm MD}\bigg[p_f \overline{P}_{{\rm out},p}^{\left({\rm c}\right)}+\left(1-p_f\right)\overline{P}_{{\rm out},p}\bigg]
\\&+p_s \left(1-P_{\rm MD}\right)\bigg[p_b \overline{P}_{{\rm out},p}^{\left({\rm c}\right)}+\left(1-p_b\right)\overline{P}_{{\rm out},p}\bigg]\Bigg)
\end{split}
\end{equation}

\subsection{Backlogged Secondary User}

Since the queues are interacting with each other, the exact analysis of the system is infeasible. In order to make the queueing analysis tractable, we make the following two approximations. We assume that a dummy data packet is transmitted from $Q_s$ when it is empty. That is, we assume a backlogged secondary user who always has data packets to transmit. In addition, we assume, as in \cite{pappas2011optimal}, that one energy packet is consumed at each time slot regardless of actual secondary activity. The first approximation reduces the primary throughput as it eliminates the possibility of an empty secondary queue, which corresponds to an interference-free situation for the PU and a reduced outage probability. The second approximation means that the energy queue empties faster than in the actual system, thereby lowering the probability that the queue is nonempty. This reduces the secondary throughput by increasing the probability that the secondary node does not have energy. Therefore, our approximations, for each value of $\lambda_p$, result in a {\bf lower bound} on the secondary service rate or throughput.


\begin{figure}
  \includegraphics[width=1\columnwidth]{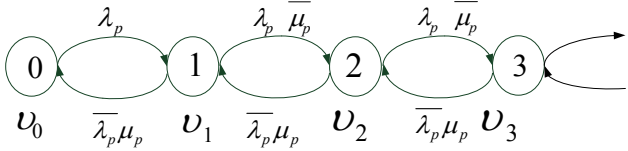}\\
  \caption{Markov chain of the PU when the secondary user does not use primary feedback. State self-transitions are omitted for visual clarity. Probabilities $\overline{\mu_p}=1-\mu_p$ and $\overline{\lambda_p}=1-\lambda_p$.}\label{fig22}
\end{figure}

Under the aforementioned assumption, the probability of the energy queue being nonempty is $\lambda_e$ \cite{pappas2011optimal}. This is because $\mu_e=1$ and the probability of the energy queue being empty is $1-\frac{\lambda_e}{\mu_e}$. The average data service rate of $Q_p$ is given by

\begin{equation}
\begin{split}
\mu_p&=\bigg(1-\lambda_e\bigg)\overline{P}_{{\rm out},p}+\lambda_e\Bigg(
\\&\left(1-p_s\right)\bigg[p_t \overline{P}_{{\rm out},p}^{\left({\rm c}\right)}+\left(1-p_t\right)\overline{P}_{{\rm out},p}\bigg]
\\&+p_s P_{\rm MD}\bigg[p_f \overline{P}_{{\rm out},p}^{\left({\rm c}\right)}+\left(1-p_f\right)\overline{P}_{{\rm out},p}\bigg]
\\&+p_s \left(1-P_{\rm MD}\right)\bigg[p_b \overline{P}_{{\rm out},p}^{\left({\rm c}\right)}+\left(1-p_b\right)\overline{P}_{{\rm out},p}\bigg]\Bigg)
\end{split}
\label{mup}
\end{equation}
\noindent Since ${\rm Pr}\{Q_p=0\}=1-\frac{\lambda_p}{\mu_p}$,
\begin{equation}
\begin{split}
\mu_s&=\lambda_e \Bigg[\bigg(1-\frac{\lambda_p}{\mu_p}\bigg) \Bigg(\left(1-p_s\right)p_t {\overline{P}}_{{\rm out},0s}+
\\&p_s p_b P_{\rm FA} {\overline{P}}_{{\rm out},1s}+p_s p_f  \left(1-P_{\rm FA}\right){\overline{P}}_{{\rm out},1s} \Bigg)+
\\&\frac{\lambda_p}{\mu_p}\Bigg(\left(1-p_s\right)p_t {\overline{P}}_{{\rm out},0s}^{\left({\rm c}\right)} +p_s p_f  P_{\rm MD} {\overline{P}}_{{\rm out},1s}^{\left({\rm c}\right)}
\\&+p_s p_b   \left(1-P_{\rm MD}\right) {\overline{P}}_{{\rm out},1s}^{\left({\rm c}\right)} \Bigg)\Bigg]
\end{split}
\label{mus}
\end{equation}

The Markov chain modeling the primary queue is provided in Fig. \ref{fig22}. Solving the state balance equations, it is straightforward to show that the probability that the primary queue has $k$ packets is
\begin{equation}
\nu_k=\nu_{\circ}\frac{1}{\overline{\mu_p}}\Bigg[\frac{\lambda_p \overline{\mu_p}}{\overline{\lambda_p }\mu_p}\Bigg]^{k}
\label{nuk}
\end{equation}
\noindent where $\overline{\lambda_p}=1-\lambda_p$ and $\overline{\mu_p}=1-\mu_p$. Using the condition $\sum_{k=0}^{\infty}\nu_k=1$,
\begin{equation}
\nu_{\circ}=1-\frac{\lambda_p}{\mu_p}
\end{equation}
\noindent For the sum $\sum_{k=0}^{\infty}\nu_k$ to exist, we should have $\lambda_p < \mu_p$. This is equivalent to Loynes' theorem which states that if the arrival and service processes of a queue are strictly stationary processes, then the queue is stable if and only if the mean arrival rate is less than the mean service rate \cite{loynes1962stability}.

\subsection{Primary User Packet Delay}
Let $D_p$ be the average delay of the primary queue. Using Little's law and (\ref{nuk}),
\begin{equation}
\begin{split}
D_p=\frac{1}{\lambda_p}\sum_{k=1}^{\infty}k\nu_k=\frac{1-\lambda_p}{\mu_p-\lambda_p}
\end{split}
\label{190990}
\end{equation}
 
\noindent For the optimal random access and sensing, we solve the following constrained optimization problem. We maximize the mean secondary service rate under the constraints that the primary queue is stable and that the primary packet delay is smaller than or equal a specified value $\overline{D_p}$. The optimization problem with
 $\mu_p$ given in (\ref{mup}) and $\mu_s$ in (\ref{mus}) can be written as
\begin{equation}
\begin{split}
& \max_{p_s,p_f,p_b,p_t} \,\,\mu_s\\
&\,\,{\rm s.t.} \,\,\,\,\  0 \le p_s,p_f,p_b,p_t \le 1\\
&\,\,\,\,\,\,\,\,\,\,\,\,\,\,\, \lambda_p\le \mu_p\\& \,\,\,\,\,\,\,\,\,\,\,\,\,\,D_p \le \overline{D_p}
\end{split}
\label{190990}
\end{equation}
This optimization problem and the others presented in this work are solved numerically\footnote{Specifically, we use Matlab's fmincon. Since the problems are nonconvex, fmincon produces a locally optimum solution. To enhance the solution and increase the likelihood of obtaining the global optimum, the program is run many times with different initializations of the optimization variables.}.

\section{Feedback-based Access}\label{sec3}
In this section we analyze the use of the primary feedback messages by the cognitive terminal. In the feedback-based access scheme, the secondary user utilizes the available primary feedback information for accessing the channel in addition to spectrum sensing. Leveraging the primary feedback is valid when it is available and unencrypted.

In the proposed scheme, the secondary user monitors the PU feedback
channel. It may overhear an acknowledgment (ACK) if the primary receiver correctly decodes the primary transmission, a negative acknowledgment (NACK) if decoding fails, or nothing if there is no primary transmission. We introduce the following modification to the protocol introduced earlier in the paper. If a NACK is overheard by the secondary user, it assumes that the PU will retransmit the lost packet during the next time slot \cite{KarimSultan}. Being sure that the PU will be active, the secondary terminal does not need to sense the channel to ascertain the state of primary activity. Therefore, it just accesses the channel with some probability $p_r$. If an ACK is observed on the feedback channel or no primary feedback is overheard, the secondary user proceeds to operate as explained earlier in Section \ref{sec2}. We assume the feedback packets are very short compared to $T$ and are always received correctly by both the primary and secondary terminals due to the use of strong channel codes. 

It is important to emphasize here the benefit of employing primary feedback. By avoiding spectrum sensing, the secondary terminal does not have to waste $\tau$ seconds for sensing. It can use the whole slot duration for transmission. As proven in Appendix C, this reduces the outage probability of the secondary link. Therefore, by differentiating between the primary states of transmission, i.e., whether they are following the reception of an ACK or not, the secondary user can potentially enhance its throughput by eliminating the need for spectrum sensing when the PU is about to retransmit a previously lost packet.

\subsection{Queueing Analysis with a Backlogged Secondary User}

\begin{figure}
  \includegraphics[width=1\columnwidth]{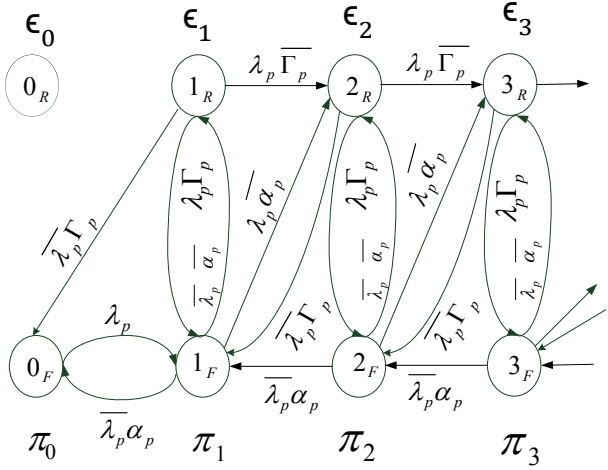}\\
  \caption{Markov chain of the PU for the feedback-based access scheme. Probabilities $\overline{\Gamma_p}=1-\Gamma_p$ and $\overline{\alpha_p}=1-\alpha_p$. State self-transitions are not depicted for visual clarity.}\label{fig2}
\end{figure}

The PU's queue evolution
Markov chain is shown in Fig. \ref{fig2}. The probability of the queue
having $k$ packets and transmitting for the first time is $\pi_k$, where
$F$ in Fig. \ref{fig2} denotes first transmission. The probability of the
queue having $k$ packets and retransmitting is $\epsilon_k$, where $R$ in
Fig. \ref{fig2} denotes retransmission. Define $\alpha_p$ as the probability of successful transmission of the PU's packet in case of first transmission and $\Gamma_p$ is the probability of successful transmission of the PU's packet in case of retransmission. It can be shown that both probabilities are given by:
 \begin{equation}\label{def}
\begin{split}
   \alpha_p&=\bigg(1-\lambda_e\bigg)\overline{P}_{{\rm out},p}+\lambda_e\Bigg(
\\&\left(1-p_s\right)\bigg[p_t \overline{P}_{{\rm out},p}^{\left({\rm c}\right)}+\left(1-p_t\right)\overline{P}_{{\rm out},p}\bigg]
\\&+p_s P_{\rm MD}\bigg[p_f \overline{P}_{{\rm out},p}^{\left({\rm c}\right)}+\left(1-p_f\right)\overline{P}_{{\rm out},p}\bigg]
\\&+p_s \left(1-P_{\rm MD}\right)\bigg[p_b \overline{P}_{{\rm out},p}^{\left({\rm c}\right)}+\left(1-p_b\right)\overline{P}_{{\rm out},p}\bigg]\Bigg)
    \end{split}
\end{equation}
\begin{equation}
\begin{split}
\Gamma_p&=\bigg(1-\lambda_e\bigg)\overline{P}_{{\rm out},p}+\lambda_e \bigg[p_r \overline{P}_{{\rm out},p}^{\left({\rm c}\right)}+\left(1-p_r\right)\overline{P}_{{\rm out},p}\bigg]
\end{split}
\end{equation}
Solving the state balance equations, we can obtain the state probabilities which are provided in Table \ref{table}. The probability $\pi_0$ is obtained using the normalization condition $\sum_{k=0}^{\infty} (\pi_k+\epsilon_k)=1$.

It should be noticed that $\lambda_p < \eta$, where $\eta$ is defined in Table \ref{table}, is a condition for the sum $\sum_{k=0}^{\infty} (\pi_k+\epsilon_k)$ to exist. This condition ensures the existence of a stationary distribution for the Markov chain and guarantees the stability of the primary queue. The service rate of the secondary user is given by:
\begin{equation}
\begin{split}
\mu_s&=\lambda_e \Bigg[\pi_0 \Bigg(\left(1-p_s\right)p_t {\overline{P}}_{{\rm out},0s}+
\\&p_s p_b P_{\rm FA} {\overline{P}}_{{\rm out},1s}+p_s p_f  \left(1-P_{\rm FA}\right){\overline{P}}_{{\rm out},1s} \Bigg)+
\\&\Bigg(\sum_{k=1}^{\infty}\pi_k\Bigg)\Bigg(\left(1-p_s\right)p_t {\overline{P}}_{{\rm out},0s}^{\left({\rm c}\right)} +p_s p_f  P_{\rm MD} {\overline{P}}_{{\rm out},1s}^{\left({\rm c}\right)}
\\&+p_s p_b   \left(1-P_{\rm MD}\right) {\overline{P}}_{{\rm out},1s}^{\left({\rm c}\right)} \Bigg)+\Bigg(\sum_{k=1}^{\infty}\epsilon_k\Bigg)p_r \overline{P}_{{\rm out},0s}^{\left({\rm c}\right)}\Bigg]
\end{split}
\label{muss}
\end{equation}
\noindent where the probability summations are given in Table \ref{table}.
\begin{table}
\renewcommand{\arraystretch}{2}
\begin{center}
\begin{tabular}{ c |c  }
    \hline\hline
    $\eta$ & $\lambda_p\alpha_p+\left(1-\lambda_p\right)\Gamma_p$ \\[5pt]\hline
  $\pi_{\circ}$ & $\frac{\eta-\lambda_p}{\Gamma_p}$ \\[5pt]\hline
  $\epsilon_{\circ}$ & 0 \\[5pt]\hline
  $\pi_1$& $\pi_{\circ}\tfrac{\lambda_p}{1-\lambda_p}\frac{\lambda_p+\left(1-\lambda_p\right)\Gamma_p}{\eta}$  \\[5pt]\hline
  $\epsilon_1$ & $\pi_{\circ}\frac{\lambda_p}{\eta}\left(1-\alpha_p\right)$ \\ [5pt]\hline
  $\pi_k,k\geq 2$ & $\pi_{\circ}\frac{\lambda_p\left(1-\alpha_p\right)}{\left(1-\eta\right)^2} \bigg[ \frac{\lambda_p \left(1-\eta\right)}{\left(1-\lambda_p\right) \eta}  \bigg]^k$ \\[5pt]\hline
  $\epsilon_k,k\geq 2$ & $\pi_{\circ}\frac{\left(1-\lambda_p\right)\left(1-\alpha_p\right)}{\left(1-\eta\right)^2} \bigg[ \frac{\lambda_p \left(1-\eta\right)}{\left(1-\lambda_p\right) \eta}  \bigg]^k$ \\[5pt]\hline
  $\sum_{k=1}^{\infty}\pi_k$ & $\pi_{\circ}\frac{\lambda_p\Gamma_p}{\eta-\lambda_p}=\lambda_p$\\[5pt]\hline
   $\sum_{k=1}^{\infty}\epsilon_k$ & $\pi_{\circ}\frac{\lambda_p}{\eta-\lambda_p}\left(1-\alpha_p\right)=\frac{\lambda_p }{\Gamma_p}\left(1-\alpha_p\right)$
\end{tabular}
\caption{State probabilities for the feedback-based access scheme.}
\label{table}
\end{center}
\end{table}

\subsection{Primary User Packet Delay}
Applying Little's law, the primary queueing delay is given by
\begin{equation}
D_p=\frac{1}{\lambda_p}\sum_{k=1}^{\infty} k \left(\pi_k+\epsilon_k\right)
\end{equation}
\noindent Using the state probabilities provided in Table \ref{table},
\begin{equation}
\begin{split}
D_p &=\frac{(\alpha_p-\eta)(\eta-\lambda_p)^2+\left(1-\lambda_p\right)^2 \left(1-\alpha_p\right) \eta}{(\eta-\lambda_p)\left(1-\lambda_p\right) \left(1-\eta\right) \Gamma_p}
  \end{split}
\end{equation}
 For a fixed $\lambda_p$, the maximum mean
service rate for the secondary user is given by solving the following
optimization problem using expression (\ref{muss}) for $\mu_s$
\begin{equation}
\begin{split}
& \max_{p_s,p_f,p_t,p_b,p_r} \,\,\mu_s\\
&\,\,{\rm s.t.} \,\,\,\,\  0 \le p_s,p_f,p_t,p_b,p_r \le 1\\
&\,\,\,\ \lambda_p \le \eta
\\
&\,\,\,\ D_p \le \overline{D_p}
\end{split}
\label{190990}
\end{equation}

\section{Numerical Results}\label{numerical}
We provide here some numerical results for the optimization problems presented in this paper and compare with a random access system without spectrum sensing. A random access scheme is simply obtained from the schemes described in this paper by setting $p_s$ to zero. We define here two variables $\delta=\frac{{\overline{P}}_{{\rm out},1s}}{{\overline{P}}_{{\rm out},0s}}$ and $\delta^c=\frac{{\overline{P}}_{{\rm out},1s}^{\left({\rm c}\right)}}{{\overline{P}}_{{\rm out},0s}^{\left({\rm c}\right)}}$, both of them are less than $1$ as shown in Appendix C. Figs.\ \ref{r1} and \ref{r9} provide a comparison between the maximum secondary rate of our proposed systems and the random access systems. The difference between the two figures is that the delay of the PU is constrained by $\overline{D_p}=2$ time slots and $\overline{D_p}=200$ time slots in Figs. \ref{r1} and \ref{r9}, respectively. The figures show clearly the advantage of employing random sensing and access over a purely random access system. Moreover, the benefit of leveraging the primary feedback is evident from the figures. Fig.\ \ref{r5} provides the optimal sensing and access probabilities corresponding to the proposed scheme employing feedback in Fig.\ \ref{r1}.

To facilitate investigating the impact of the specified delay constraint, Fig.\ \ref{delay_fig} provides the effect of varying this constraint on the secondary service rate. As is clear from the figure, the secondary service rate is reduced when the primary queueing delay constraint is more strict. Fig.\ \ref{rx} shows the impact of the energy arrival rate. As expected, the secondary service rate is reduced when $\lambda_e$ decreases. Finally, Fig.\ \ref{ry} shows the impact of the MPR capability. Without MPR capability, collisions are assumed to lead to sure packet loss. Therefore, a collision model without MPR corresponds to the case of the probabilities of correct reception being zero when there are simultaneous transmissions. As shown in  Fig.\ \ref{ry}, the secondary service rate is reduced when there is no MPR capability.


\section{conclusion} \label{secfinal}
In this paper, we discussed an energy harvesting cognitive radio. We proposed a novel sensing/access scheme to enable the secondary user efficiently use its own energy to access and sense the channel. We considered two schemes based on feedback of the primary receiver channel. We compared our results with the random access scheme with and without using the feedback information. The results have shown the benefit of employing the primary feedback in making secondary access decisions.

\section*{Appendix A}
We adopt a flat fading channel model and assume that the channel gains remain constant over the duration of the time slot. We do not assume the availability of transmit channel state information (CSI) at the transmitting terminals. The transmitters adjust their transmission rates depending on when they start transmission during the time slot.
Assuming that the number of bits in a packet is $b$, the transmission rate is
\begin{equation}
r_i=\frac{b}{T\left(1-i\frac{\tau}{T}\right)}.
\label{r_i}
\end{equation}
\noindent where $\tau$ is the amount of time spent to sense the channel, $\tau < T$, $i=0$ if transmission proceeds at the very beginning of the time slot, and $i=1$ if transmission is preceded by a spectrum sensing period of $\tau$ units of time. Outage for a secondary user occurs when the transmission rate exceeds the channel capacity
\begin{equation}
P_{{\rm out},is}={\rm Pr}\{r_i > W \log_{2}\left(1+\gamma \beta\right)\}
\end{equation}
\noindent where $W$ is the bandwidth of the channel, $\gamma$ is the received SNR when the channel gain is equal to unity, and $\beta$ is the channel gain, which is exponentially distributed in the case of Rayleigh fading.

The outage probability can be written as
\begin{equation}
P_{{\rm out},is}={\rm Pr}\Big\{\beta<\frac{2^{\frac{r_i}{W}}-1}{\gamma}\Big\}
\end{equation}
\noindent Assuming that the mean value of $\beta$ is $\overline{\beta}$,
\begin{equation}
P_{{\rm out},is}=1-\exp\bigg(-\frac{2^{\frac{r_i}{W}}-1}{\gamma\overline{\beta}}\bigg)
\end{equation}
\noindent Let $\overline{P}_{{\rm out},is}=1-P_{{\rm out},is}$ be the probability of correct reception. It is therefore given by
\begin{equation}\label{correctreception}
\overline{P}_{{\rm out},is}=\exp\bigg(-\frac{2^{\frac{b}{TW\left(1-i\frac{\tau}{T}\right)}}-1}{\gamma\overline{\beta}}\bigg)
\end{equation}
\noindent The probability $\overline{P}_{{\rm out},p}$ for the PU is given by a similar expression with $i=0$ as the PU transmits starting from the beginning of the time slot and the relevant primary link parameters.

\begin{figure}
  \includegraphics[width=1\columnwidth]{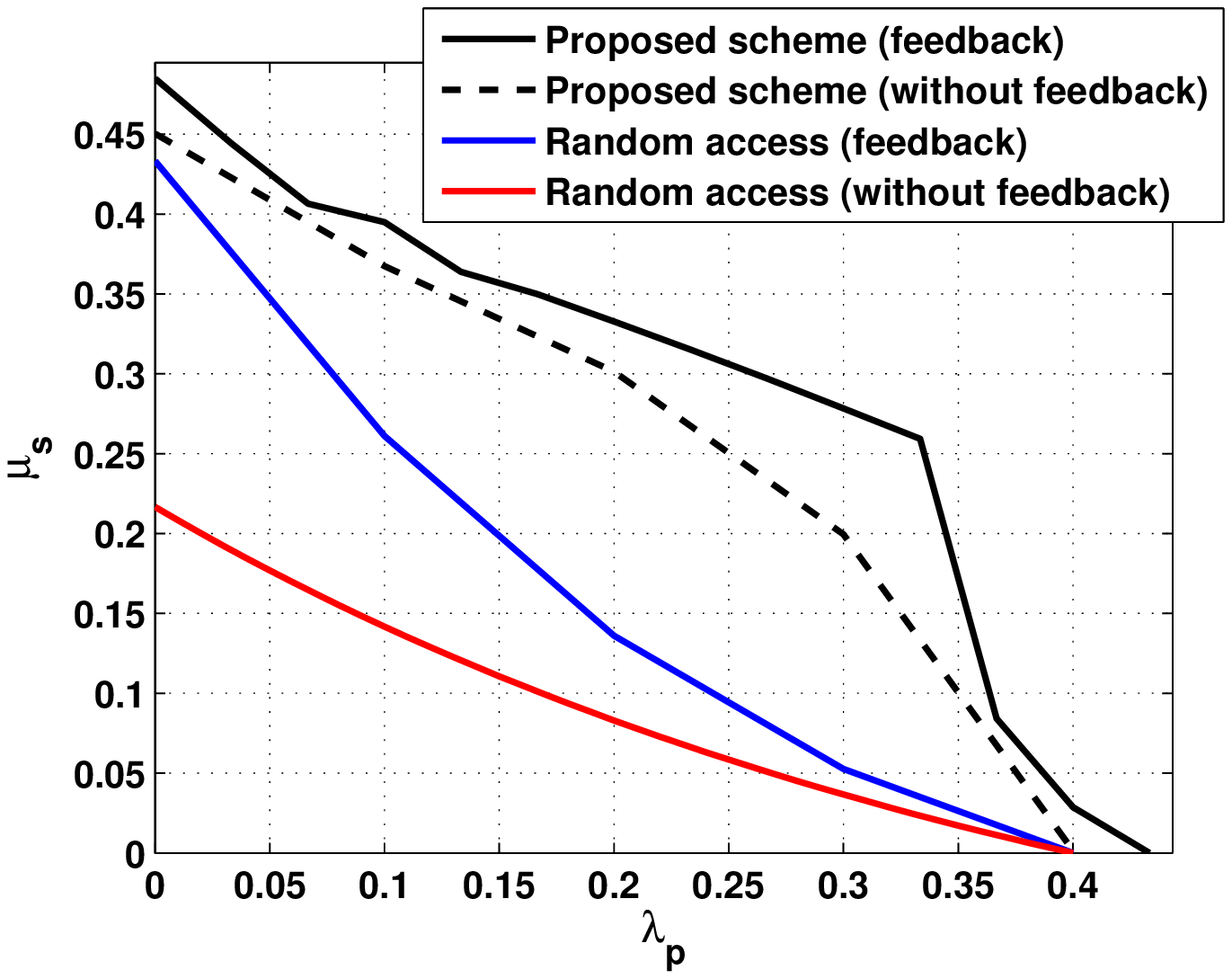}\\
   \caption{Maximum secondary service rate for the parameters: $\lambda_e=0.8$, $\overline{P}_{{\rm out},p}=0.7$, ${\overline{P}}_{{\rm out},p}^{\left({\rm c}\right)}=0.14$, $\overline{P}_{{\rm out},0s}=0.6065$, ${\overline{P}}_{{\rm out},0s}^{\left({\rm c}\right)}=0.1820$, $\delta=0.9782$,  $\delta^{\rm c}=0.8$, $P_{\rm FA}=0.1$, $P_{\rm MD}=0.08$, and $\overline{D_p}=2$ time slot.}\label{r1}
\end{figure}
\begin{figure}
  \includegraphics[width=1\columnwidth]{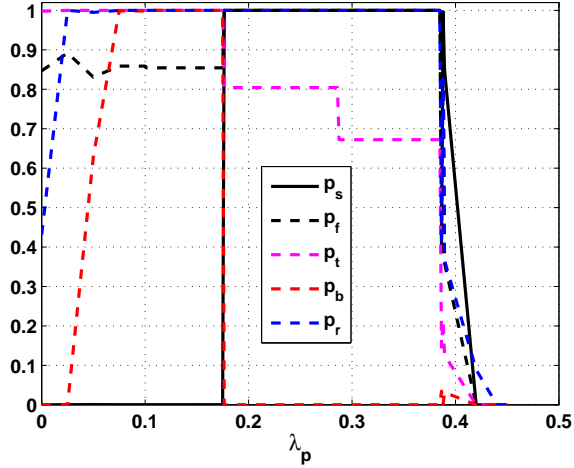}\\
   \caption{Optimal values of the sensing and access probabilities for the feedback-based system depicted in Fig.\ \ref{r1}.}\label{r5}
\end{figure}

\begin{figure}
  \includegraphics[width=1\columnwidth]{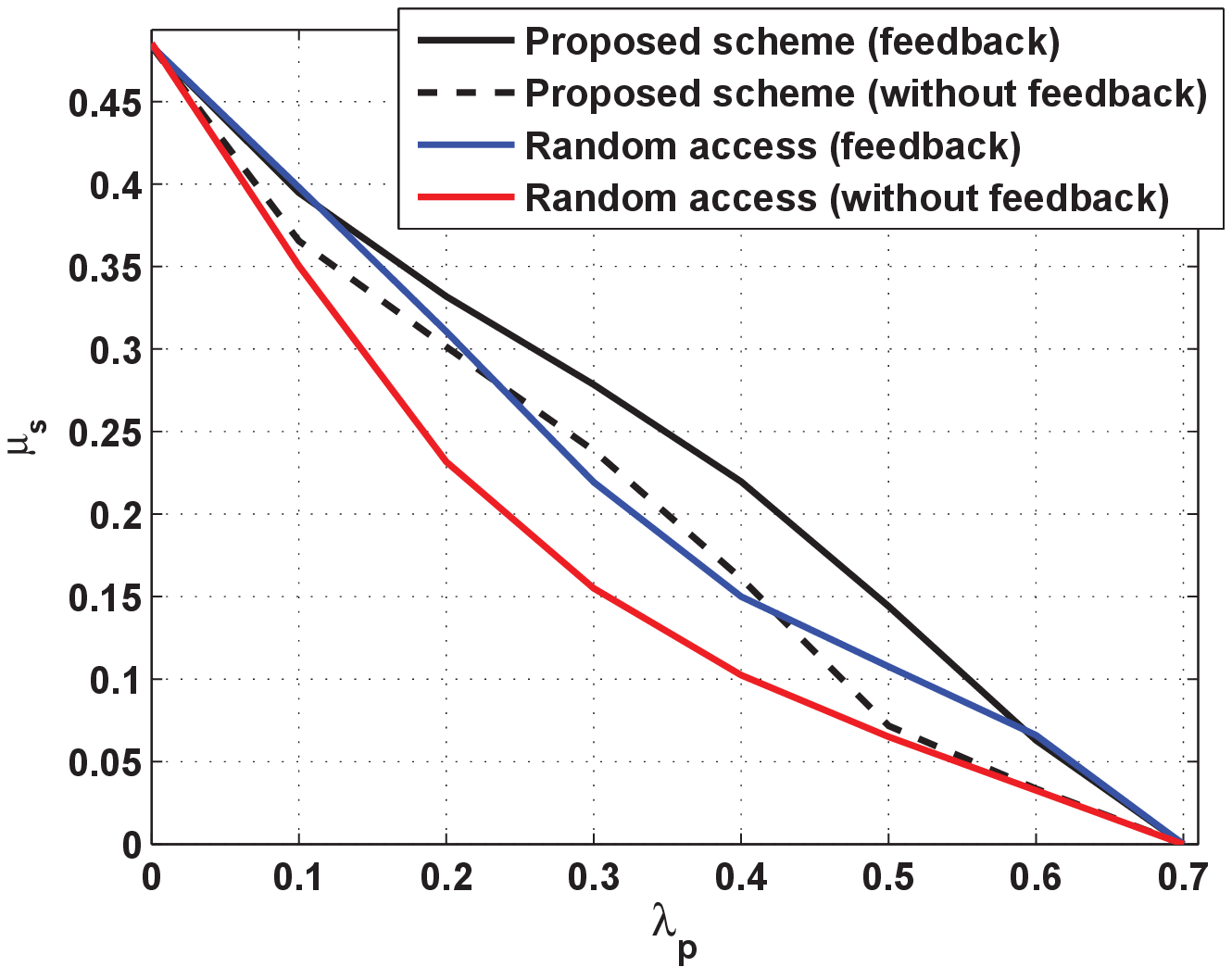}\\
   \caption{Maximum secondary service rate for the parameters: $\lambda_e=0.8$, $\overline{P}_{{\rm out},p}=0.7$, ${\overline{P}}_{{\rm out},p}^{\left({\rm c}\right)}=0.14$, $\overline{P}_{{\rm out},0s}=0.6065$, ${\overline{P}}_{{\rm out},0s}^{\left({\rm c}\right)}=0.1820$, $\delta=0.9782$,  $\delta^{\rm c}=0.8$, $P_{\rm FA}=0.1$, $P_{\rm MD}=0.08$, and $\overline{D_p}=200$ time slot.}\label{r9}
\end{figure}

\begin{figure}
  \includegraphics[width=1\columnwidth]{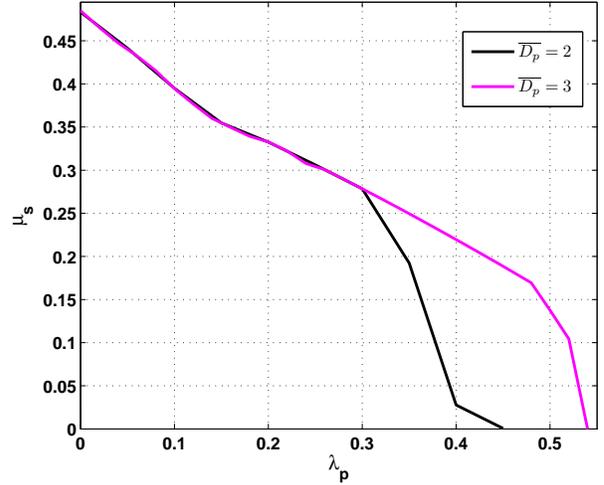}\\
   \caption{Maximum secondary service rate for different values of the primary packet delay constraint. Parameters used to generate the figure are: $\lambda_e=0.8$, $\overline{P}_{{\rm out},p}=0.7$, ${\overline{P}}_{{\rm out},p}^{\left({\rm c}\right)}=0.14$, $\overline{P}_{{\rm out},0s}=0.6065$, ${\overline{P}}_{{\rm out},0s}^{\left({\rm c}\right)}=0.1820$, $\delta=0.9782$,  $\delta^{\rm c}=0.8$, $P_{\rm FA}=0.1$, and $P_{\rm MD}=0.08$}\label{delay_fig}
\end{figure}

\begin{figure}
  \includegraphics[width=1\columnwidth]{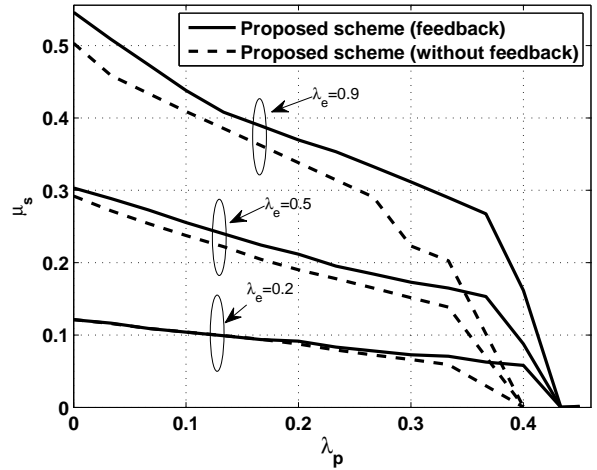}\\
   \caption{Impact of the energy arrival rate $\lambda_e$ on the maximum secondary service rate for the parameters: $\overline{P}_{{\rm out},p}=0.7$, ${\overline{P}}_{{\rm out},p}^{\left({\rm c}\right)}=0.14$, $\overline{P}_{{\rm out},0s}=0.6065$, ${\overline{P}}_{{\rm out},0s}^{\left({\rm c}\right)}=0.1820$, $\delta=0.9782$,  $\delta^{\rm c}=0.8$, $P_{\rm FA}=0.1$, $P_{\rm MD}=0.08$, and $\overline{D_p}=2$ time slot.}\label{rx}
\end{figure}

\begin{figure}
  \includegraphics[width=1.09\columnwidth]{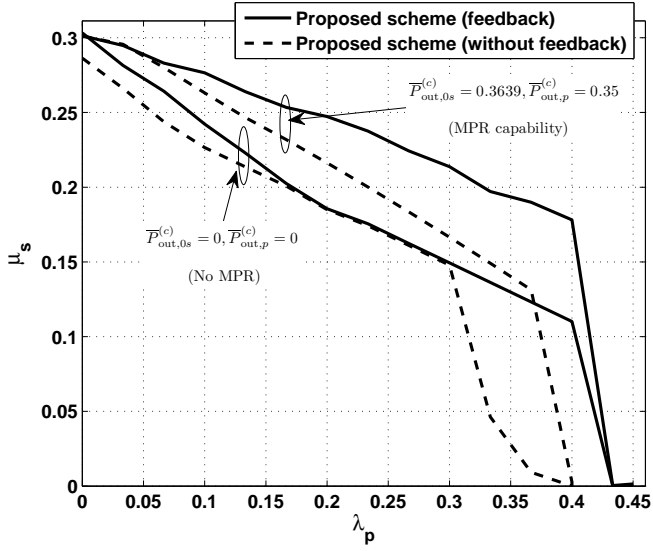}\\
   \caption{Impact of the MPR capability on the maximum secondary service rate for the parameters: $\lambda_e=0.5$, $\overline{P}_{{\rm out},p}=0.7$, $\overline{P}_{{\rm out},0s}=0.6065$, $\delta=0.9782$,  $\delta^{\rm c}=0.8$, $P_{\rm FA}=0.1$, $P_{\rm MD}=0.08$, and $\overline{D_p}=2$ time slot.}\label{ry}
\end{figure}

\section*{Appendix B}
We derive here a generic expression for the outage probability at the receiver of link $j$ when there is concurrent transmission from the transmitter of link $v$. Outage occurs when the transmission rate $r_i$, given by (\ref{r_i}), exceeds the channel capacity
\begin{equation}
P_{{\rm out},ij}^{\left({\rm c}\right)}={\rm Pr}\biggr\{r_i > W \log_{2}\left(1+\frac{\gamma_{j} \beta_{j}}{\gamma_{v} \beta_{v}+1}\right)\biggr\}
\end{equation}
\noindent where the superscript c denotes concurrent transmission, $\gamma_j$ is the received SNR at the receiver of link $j$ without interference when the channel gain $\beta_j$ is equal to unity, and $\gamma_v$ is the received SNR at the receiver of link $j$ when it only receives a signal from the interfering transmitter of link $v$ given that the gain of the interference channel, $\beta_v$, is equal to unity.
The outage probability can be written as
\begin{equation}\label {1900}
P_{{\rm out},ij}^{\left({\rm c}\right)}={\rm Pr}\Big\{\frac{\gamma_{j} \beta_{j}}{\gamma_{v} \beta_{v}+1}<{2^{\frac{r_i}{W}}-1}\Big\}
\end{equation}
\noindent Assuming that $\beta_j$ and $\beta_v$ are exponentially distributed with means $\overline{\beta_{j}}$ and $\overline{\beta_{v}}$, respectively, we can use the probability density functions of these two random variables to obtain the outage as
 \begin{eqnarray*}\label{193}
 P_{{\rm out},ij}^{\left({\rm c}\right)}=1-\frac{1}{1+\Big( {2^{\frac{r_i}{W}}-1} \Big)\frac{\gamma_v\overline{\beta_v}}{ \gamma_j\overline{\beta_j}}} {e^{-\frac{{2^{\frac{r_i}{W}}-1}}{\gamma_j \overline{\beta_j}}}}
\end{eqnarray*}
\noindent The probability of correct reception $  \overline{P}^{\left({\rm c}\right)}_{{\rm out},ij}=1-P^{\left({\rm c}\right)}_{{\rm out},ij}$ is thus given by

  \begin{eqnarray}\label{conctra}
 \overline{P}_{{\rm out},ij}^{\left({\rm c}\right)}=\frac{\overline{P}_{{\rm out},ij}}{1+\Big({2^{\frac{b}{TW\left(1-\frac{i\tau}{T}\right)}}-1} \Big)\frac{\gamma_v\overline{ \beta_v}}{\gamma_j \overline{\beta_j}}}
\end{eqnarray}
\noindent As is obvious, the probability of correct reception is lowered in the case of interference.
\section*{Appendix C}
We prove here that $\overline{P}_{{\rm out},0j} > \overline{P}_{{\rm out},1j}$. As a function of $\tau$, $\overline{P}_{{\rm out},1j}$ is given by (\ref{correctreception}) with $i=1$. Let $x=1-\frac{\tau}{T}$ where
$x \in [0,1]$. Assuming that the energy unit used per slot is $e$, the transmit power is $\frac{e}{T-\tau}=\frac{e}{Tx}$. This means that the received SNR $\gamma$ is inversely proportional to $x$. The exponent in (\ref{correctreception}) with $i=1$ is thus proportional to $\boldsymbol{g(x)=x (\exp(\frac{a}{x})-1)}$ where $a =\frac{b \ln 2} {WT} > 0$. Differentiating $g(x)$ with respect to $x$, the derivative is
\begin{equation}
\begin{split}
-1+\Big[1-\frac{a}{x}\Big] \exp(\frac{a}{x})&=-1+\Big[1-\frac{a}{x}\Big]\sum_{k=0}^{\infty}\frac{1}{k!}\Big(\frac{a}{x}\Big)^{k}\\&=
-1+\!\sum_{k=0}^{\infty}\frac{1}{k!}\Big(\frac{a}{x}\Big)^{k}\!-\!\sum_{k=0}^{\infty}\frac{1}{k!}\Big(\frac{a}{x}\Big)^{k+1}\\&=
\sum_{k=1}^{\infty}\frac{1}{k!}\Big(\frac{a}{x}\Big)^{k}-\sum_{k=1}^{\infty}\frac{1}{(k-1)!}\Big(\frac{a}{x}\Big)^{k}\\&=
\sum_{k=1}^{\infty}\bigg[\frac{1}{k!}-\frac{1}{(k-1)!}\bigg]\Big(\frac{a}{x}\Big)^{k+1}<0
\end{split}
\end{equation}
Therefore, the derivative is always negative. Since $x=1-\frac{\tau}{T}$, function $g(x)$ increases with $\tau$. This means that $\overline{P}_{{\rm out},1j}$ decreases with $\tau$ and its maximum value occurs when the transmission starts at the beginning of the time slot, i.e., $\tau=0$. This proves that $\overline{P}_{{\rm out},0j} > \overline{P}_{{\rm out},1j}$ for $\tau > 0$. For the case of concurrent transmission given in (\ref{conctra}), the numerator is $\overline{P}_{{\rm out},1j}$, which decreases with $\tau$. The denominator is proportional to $g(x)$ which has been shown to decrease with $x$ and, hence, increase with $\tau$. This means that $\overline{P}^{\left({\rm c}\right)}_{{\rm out},1j}$ decreases with $\tau$, i.e.,  $\overline{P}^{\left({\rm c}\right)}_{{\rm out},0j} > \overline{P}^{\left({\rm c}\right)}_{{\rm out},1j}$.

\bibliographystyle{IEEEtran}
\bibliography{IEEEabrv,energy_bib}
\end{document}